\documentclass[twocolumn,floatfix,prb,eqsecnum,showpacs]{revtex4}
\usepackage{graphicx}
\usepackage{amsmath}
\usepackage{bm}
\begin{document}

\title{Fermi edge singularity and finite frequency spectral features in a semi-infinite 1D wire}

\author{A. Sheikhan and I. Snyman}
\email{isnyman@sun.ac.za}
\affiliation{National Institute for Theoretical Physics, Private Bag X1, 7602 Matieland, South Africa}
\date{March 2012}
\begin{abstract} 
We theoretically study a charge qubit interacting with electrons in a semi-infinite 1D wire. The system displays
the physics of the Fermi edge singularity. Our results generalize known results for the Fermi-edge system to
the regime where excitations induced by the qubit can resolve the spatial structure of the scattering region. 
We find resonant features in the qubit tunneling rate as a function of the qubit level splitting. They occur at integer
multiples of $h v_F/l$. Here $v_F$ is the Fermi velocity of the electrons in the wire, and $l$ is the
distance from the tip of the wire to the point where it interacts with the qubit. These features are due to
a single coherent charge fluctuation in the electron gas, 
with a half-wavelength that fits into $l$ an integer number of times.    
As the coupling between the qubit and the wire is increased, the resonances are washed out. This is
a clear signature of the increasingly violent Fermi-sea shake-up that accompanies strong coupling.
  
\end{abstract}
\maketitle

\section{Introduction}
Systems in which a localized impurity, with an internal quantum mechanical degree of freedom, interacts
with an electron gas, play an important role in many-body theory.  On the one hand, they allow theorists to 
investigate interaction effects and many-body correlations beyond the perturbative regime.
On the other hand they explain observed phenomena such as the resistance minimum (as a function of
temperature) in dilute magnetic alloys, i.e. the Kondo effect.\cite{kon64}   
Another impurity phenomenon that has been studied extensively is the so called Fermi edge
singularity.\cite{mah67,noz69} In its original incarnation, the effect refers to power law singularities 
in the soft-x-ray absorption, emission, and photoemmision spectra of metallic samples.
As with the Kondo effect,\cite{pus04} the phenomenon has received renewed attention due to 
technological breakthroughs in nano-physics and quantum transport. The same physics
that is behind the Fermi edge singularity describes for instance tunneling into
and out of a small quantum dot coupled to an electron reservoir.\cite{aba04} Recent studies have 
focused on non-equilibrium,\cite{muz03,amb04,aba05,sny07,gut10,bet11,gut11} 
non-stationary,\cite{bet11b} and band structure\cite{mkh11} phenomena.

In this paper, we confine our attention to the equilibrium situation but consider
a setup in which the spectral function
has interesting features at energies away from the singularity.        
The setup can be realized with current technology.
The system we study consists of an electron gas interacting with 
a two level system (charge qubit). A concrete realization of the qubit
could be an electron that may occupy the lowest two states of a double quantum dot.\cite{elz03,pet04}  
From the point 
of view of the electron gas, the qubit acts as a dynamic localized impurity, while from the
point of view of the qubit, the electron gas acts as a dissipative environment. 
The qubit state-space is spanned by the vectors
$\left|+\right>$ and $\left|-\right>$. The total Hamiltonian for the system is $H=H_0+H_T$ where
\begin{eqnarray}
H_0&=&(H_++\varepsilon)\left|+\right>\left<+\right|+H_-\left|-\right>\left<-\right|\nonumber\\
H_T&=&\gamma\left|+\right>\left<-\right|+\gamma^*\left|-\right>\left<+\right|.\label{h1}
\end{eqnarray}
The energy $\varepsilon$ represents a gate voltage that controls the qubit level splitting and $\gamma$ is 
a small tunneling amplitude between the two qubit states. Both these parameters are typically under 
experimental control.

The Hamiltonians $H_\pm=T+V_\pm$ describe the electron gas. The kinetic term $T$ is the
same for both Hamiltonians. The finite range potentials $V_\pm$ represent the electrostatic potential 
produced by the qubit. This potential depends on the internal state of the qubit, 
so that $V_+\not=V_-$.  We will 
specify the system in more detail in Sec.\,\ref{sys}. 
  
The quantity of interest in this article is the qubit tunneling rate $W$, which is defined as follows:
Consider the situation where initially the tunneling amplitude $\gamma$ is zero, and the qubit
is prepared in the state $\left|+\right>$. The electron gas is allowed to equilibrate. We assume zero temperature
so that it equilibrates to the Fermi-sea ground state $\left|F+\right>$
of the Hamiltonian $H_+$. At time $t=0$,  
the tunneling amplitude $\gamma$ is then switched on,
and the state $\left|+\right>$ acquires a finite lifetime.
Provided that this lifetime is long enough that we can still speak of well-defined qubit levels,
the probability $n_+(t)$ to find the qubit in the 
state $\left|+\right>$ decays exponentially\cite{f1}
\begin{equation}
n_+(t)=e^{-Wt},\label{exp}
\end{equation} 
where $W$ is the qubit tunneling rate.

In order to  formulate a quantitative criterion 
for when exponential decay occurs, it is useful to     
define an energy 
\begin{equation}
\omega=\varepsilon+E_0^{(+)}-E_0^{(-)}.\label{om}
\end{equation}
Here $E_0^{(\pm)}$ are the ground state energies associated with the Fermi sea ground states
$\left|F_\pm\right>$ of $H_\pm$. For the system described by $H_0$, the minimum
energy difference between a configuration (of qubit plus Fermi gas) with the qubit in the state $\left|+\right>$ 
and one with the qubit in the state $\left|-\right>$
is $\omega$.
Due to the finite lifetime $W^{-1}$, there is an uncertainty of order $W$ in the 
qubit level splitting. For the energy levels associated with $\left|\pm\right>$ to
remain well-defined, the minimum energy difference $\omega$ between a $\left|+\right>$ and a $\left|-\right>$
configuration must be much larger than this uncertainty, i.e.
\begin{equation}
\omega\gg W.\label{crit}
\end{equation}
In this regime $n_+(t)$ shows exponential decay. 

By applying Fermi's Golden, one obtains
\begin{equation}
W=|\gamma|^2\int_{-\infty}^\infty dt\,e^{i\varepsilon t}\left<F+\right|e^{iH_+t}e^{-iH_-t}\left|F+\right>.\label{w1}
\end{equation}
(See Appendix \ref{appa} for details). We will use this result to obtain an explicit expression for $W$
in terms of the energy $\varepsilon$ and the potentials $V_\pm$. In the original incarnation of the problem,
this quantity corresponds to the photoemmision spectrum, i.e. the intensity of electrons ejected from
the metal, at a fixed x-ray frequency $\varepsilon$. (See for instance Sec. IV of Ref. \onlinecite{oth90}.)

In the language of the Fermi edge singularity, the quantity
$\left<F+\right|e^{iH_+t}e^{-iH_-t}\left|F+\right>$, that appears in Eq.\,\ref{w1} 
is known as the closed loop factor. (See for instance Sec. III D of Ref. \onlinecite{oth90}.) 
It is known that, for $t$ much larger 
than the time an individual electron spends in the scattering region
\begin{equation}
\left<F+\right|e^{iH_+t}e^{-iH_-t}\left|F+\right>\simeq e^{-i\Delta E t}\left(i\Lambda t\right)^{-\alpha},\label{cl1}
\end{equation}
where $\Delta E=E_0^{(-)}-E_0^{(+)}$ is the difference between the ground state energies of $H_-$ and $H_+$ and 
$\Lambda$ is an ultra-violet energy scale. (The branch with
$\arg(i\Lambda t)=\pm \pi/2$ is implied.) The power law exponent $\alpha$ is determined by the 
single particle scattering matrices $S_\pm$ associated with the fermion Hamiltonians. Explicitly\cite{aba04,amb04} 
\begin{equation}
\alpha={\rm Tr}\left[\left(\frac{\ln S_+ S_-^\dagger}{2\pi}\right)^2\right].\label{alpha0}
\end{equation}
For sufficiently small $\omega$, the asymptotic form of Eq.\,\ref{cl1} 
gives rise to a tunneling rate
\begin{equation}
W=\frac{2\pi|\gamma|^2}{\Gamma(\alpha)}\left(\frac{\omega}{\Lambda}\right)^\alpha\frac{1}{\omega}\theta(\omega),
\label{w}
\end{equation}
with $\theta(\omega)$ the unit step function, i.e. $\theta(\omega)=1$ for $\omega>0$ and $\theta(\omega)=0$
for $\omega<0$. 
This power law remains valid while $\omega\ll\min\{v_F/l,E_F,D-E_F\}$. Here $v_F$ is the Fermi velocity,
$E_F$ is the Fermi energy measured from the bottom of the conduction band, and $D$ is the band width. The
length scale $l$ is the size of the scattering region. This 
is not necessarily the same length scale as the range
of the qubit interaction potentials $V_\pm$, which we denote $a$. Consider for instance a qubit
placed next to a semi-infinite wire. (See Fig.\,\ref{f0}.) Here the size of the scattering region is the distance from
the tip of the wire to the point closest to the qubit, which can be much larger that the range
of the potential produced by the charge of the qubit. In general $l\geq a$.

The condition $\omega\ll v_F/l$
comes about as follows.  
An energy $\omega$ corresponds to density fluctuations in the electron gas with wavelengths
at least $v_F/\omega$. 
The result of Eq.\,\ref{w} breaks down as soon as this wavelength is short enough for
these excitations to resolve the spatial structure of the scattering region.
The restriction $\omega\ll \min\{E_F,\, D-E_F\}$ is due to the fact that Eq.\,\ref{w} becomes
invalid when particle or hole excitations are created in the electron gas close to the band edges.  

An analytical expression for $\Lambda$ was obtained by Tanabe and Othaka\cite{tan85} 
in the regime $E_F\ll v_F/l$,
where the wavelength of electrons near the Fermi level are too long to resolve the spatial structure
of the scattering region, so that the potentials $V_\pm$ may be approximated as $\delta$-functions. 
(This is referred to as the limit of contact potentials.) $\Lambda$ was found to be of the order of $E_F$.
In the same limit, approximate results for the finite $\omega$ behavior of
$W$ has been obtained. As a function of $\omega$, these results contain features on the 
scale of $E_F$ that are associated with the
band structure of the model. For more detail the reader is referred to the review [\onlinecite{oth90}].

The regime of $E_F\ll v_F/l$ applies in a semi-conductor, where it is not 
uncommon for the Fermi wave-length to be large compared
other length scales in the problem. However, the opposite
regime, where $E_F\gg v_F/l$, also has physical relevance: In a metallic sample, the Fermi wavelength is
comparable to the lattice constant, while all length-scales associated with the potential 
are much larger. We are not aware of any work in which this regime is investigated. 

In this article we study the regime where $E_F\gg v_F/a\geq v_F/l$.
There are two significant differences between this regime and the previously studied regime.
Firstly, the ultra-violet energy $\Lambda$ is no longer of order $E_F$, but rather is determined by the
potential $v_F$. Secondly, the rate $W$ as function of $\omega$ starts deviating from the
power law of Eq.\,\ref{w}, at energies $\sim v_F/l$ rather than at energies $\sim E_F$. The
source of the deviations is no longer related to band structure, but to excitations resolving
the spatial structure of the scattering region. We confine our attention to
the case of an electron gas in a single chiral channel at zero-temperature. We pay particular
attention to the example mentioned above of a qubit interacting with a semi-infinite wire, where
$l\gg a$. We were able to obtain exact analytical expressions for $\Lambda$ and for the closed loop factor at
arbitrary times. We find that $\Lambda<v_F/a$, and that $\Lambda$ depends only on the shape, not the 
magnitude, of $V_\pm$, i.e. scaling $V_\pm\to c\,V_\pm$ leaves $\Lambda$ unchanged.
We were able to compute the tunneling rate $W(\omega)$ away from the threshold $\omega\to0^+$.
We find that $W(\omega)$ has resonant features at an energy scale $v_F/l$, 
that reveal the nature of many-body correlations induced by the qubit. Our main results are contained in
Eqs.\,\ref{qt},  \ref{lambda1}, \ref{lam}, and \ref{Wl2}.  

Our analysis is based on the approximation of 
taking the limit $E_F\to\infty$ and linearizing the electrons' dispersion relation around the 
Fermi level, while still taking into account the full spatial dependence of the potentials $V_\pm$. 
As pointed out by Gutman et al. (footnote 36 of Ref.\,\onlinecite{gut10}) special care must be taken
when linearizing the dispersion relation in order to account for the anomalous contribution to
$\left<F+\right|e^{iH_+t}e^{-iH_-t}\left|F+\right>$ that is related to the Schwinger anomaly.\cite{sch59} 
In the derivation that we present, this anomalous contribution appears quite naturally.\cite{mat65} 
Our results are obtained by means of bosonization,\cite{hal81,del98} the application of which to the Fermi
edge singularity was pioneered by Schotte and Schotte.\cite{sch69} Bosonization maps 
the Hamiltonian of Eq.\,\ref{h1} onto an equivalent spin-boson model,\cite{leg87} where a spin is coupled 
to a bosonic bath. For the example of a semi-infinite wire interacting with a qubit at a point on
the wire that is a distance $l\gg a$ from the tip of the wire, the bosonic bath spectrum has
non-trivial structure. This in turn is what leads to the non-trivial finite $\omega$ behavior of
the tunneling rate $W(\omega)$.

The rest of this article is structured as follows. In Sec.\,\ref{sys} we specialize
to a Fermi gas consisting of a single chiral channel, and introduce a model
to describe a semi-infinite wire interacting with a qubit at a point a distance
$l$ from the tip of the wire. 
In Sec.\,\ref{sec4} we collect the results from the theory bosonization that are
required for our analysis. We also discuss Anderson's orthogonality catastrophe from the point
of view provided by bosonization. In Sec.\,\ref{sec5} we give a general and exact 
formula for the closed loop factor. This allows us to calculate the ultraviolet energy scale $\Lambda$
exactly. In Sec.\,\ref{sec6} we apply the general results of 
Sec.\,\ref{sec5} to the specific system introduced in Sec.\,\ref{sys}, 
for which the tunneling rate $W(\omega)$ has
non-trivial features at finite $\omega$.   

\section{a Single chiral channel}
\label{sys}
  
Here we give a mathematical definition of the type of electron gas we study.
Associated with the electrons in a chiral channel of length $L$ with periodic boundary conditions 
are  creation and annihilation operators $\psi^\dagger(x)$ and $\psi(x)$ 
that respectively create or annihilate a fermion in the state $\left|x\right>$ localized 
at position $x$. They obey the usual anti-commutation relations
\begin{equation}
\{\psi(x),\psi^\dagger(x')\}=\sum_{n=-\infty}^\infty\delta(x-x'-nL),
\end{equation}
and are periodic with period $L$. At the point in our derivation where it becomes
convenient to do so, we send the system size $L$ to infinity.

The non-interacting many-fermion Hamiltonians $H_\pm$ have the same linear dispersion
but different external potentials. Without loss of generality (see Appendix \ref{appaa} for details), 
we can set the external potential in $H_+$ to zero, so that  
\begin{equation}
H_+=\int_{-L/2}^{L/2}dx\,\psi^\dagger(x)\left(-i\partial_x-\mu\right)\psi(x),\label{disp}
\end{equation}
while $H_-=H_++V$ with   
\begin{equation}
V=\int_{-L/2}^{L/2}dx\, v(x)\rho(x),\hspace{5mm}\rho(x)=\psi^\dagger(x)\psi(x)\label{pot}
\end{equation}
(Here we work in units where the Fermi velocity $v_F=1$.)
Associated with $H_-$ is the one-dimensional scattering matrix $e^{-iv_0}$ where
\begin{equation}
v_0=\int_{-L/2}^{L/2}dx\,v(x).\label{v0}
\end{equation}
The ground state
of $H_+$ is the Fermi-sea 
\begin{equation}
\left|F+\right>=\prod_{k\leq \mu} c_k^\dagger\left|0\right>,
\end{equation}
where $\left|0\right>$ is the state with no particles, the operator
\begin{equation}
c_k^\dagger=\frac{1}{\sqrt{L}}\int_{-L/2}^{L/2}dx\,e^{ikx}\psi^\dagger(x),
\end{equation}
creates a fermion in a momentum eigenstate and $k$ is quantized in integer multiples of 
$2\pi/L$.

\begin{figure}[tbh]
\begin{center}
(a)\includegraphics[width=.9\columnwidth]{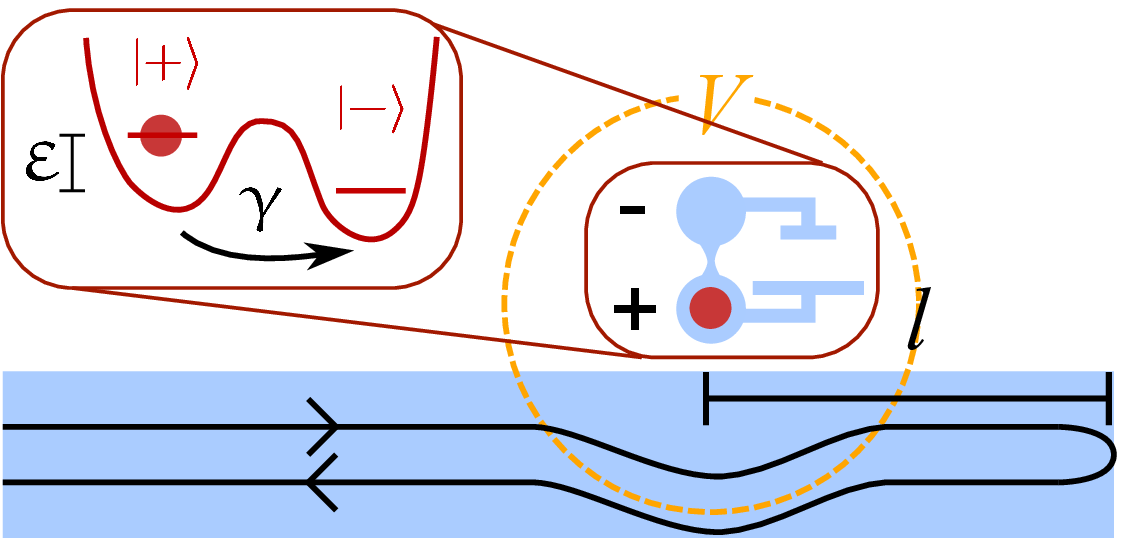}
(b)\includegraphics[width=.9\columnwidth]{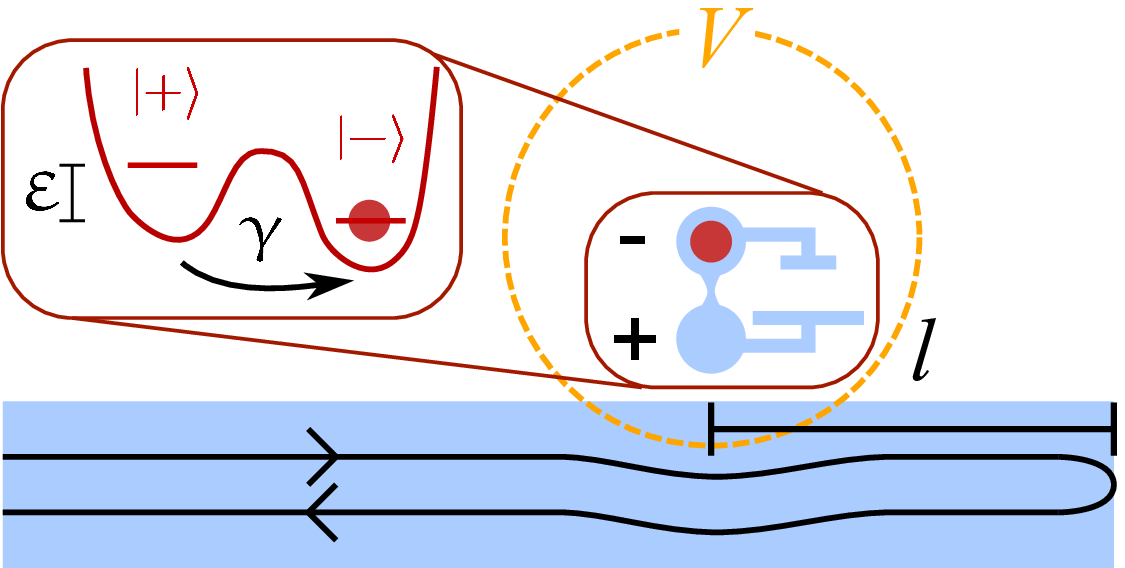}
\caption{The system described by the potential $v(x)$ of Eq.\,\ref{sinf}. (a) shows the system
with the qubit in the initial state $\left|+\right>$ and (b) shows the system after the transition.
The shaded rectangle represents the semi-infinite wire. The U-shaped contour indicates the single
chiral channel in which electrons propagate. 
The diagram is only schematic. In an actual realization, the left and right propagating electrons
need not be spatially separated, i.e. the U-shape of the contour may be squashed into a line. 
The distance $l$ between the tip of the wire and the
point on the wire closest to the qubit is indicated.   
The qubit is represented as a double quantum
dot with a single electron in it. The dashed circle indicates the range of the potential through 
which the qubit and the wire interact. In the exploded view, the qubit level spacing $\varepsilon$ and
the tunneling amplitude $\gamma$ are indicated. While the range of the potential and the distance
$l$ are of similar size in the figure, we will investigate the regime where $l$ is much larger than
the range of the potential in the text.
\label{f0}}
\end{center}
\end{figure}

The general results we obtain
will be applied to the case where the electron gas resides in a semi-infinite 1D quantum wire.
In the limit of a large Fermi energy, this system is mapped onto Eqs.\,\ref{h1}, \ref{disp}, and \ref{pot}
through the standard trick of ``unfolding'', so that coordinates $x$ and $-x$ refer to the same spatial
point, but to ``different sides of the road'', i.e.  $\psi^\dagger(-x)$ and $\psi^\dagger(x)$ create
electrons at the same position but moving in opposite directions.\cite{fab95} Thus the potential $v(x)$
is symmetric about $x=0$. The system is depicted in Figure \ref{f0}.
As a simple model for the interaction between 
the qubit and the electron gas, we will take $v(x)=u(x-l)+u(x+l)$ with 
\begin{equation}
u(x)=\left(\frac{a v_0}{2\pi}\right)\frac{1}{x^2+a^2}.\label{sinf}
\end{equation}
Here $l$ is the distance from the tip of the wire to the point on the wire nearest to the qubit and
$a$ is the range of the potential produced by the qubit. 
This choice of $u(x)$ allows us to obtain an exact analytical expression for $W(\omega)$.
In the regime where $l\gg a$ we expect qualitatively similar results for any choice of $u(x)$ that
is localized to a region of length $\sim a$.


\section{Bosonization}
\label{sec4}
In this section we collect together the known operator bosonization 
results that are required for our analysis. For a tutorial derivation, we refer the reader to Ref.\,\onlinecite{del98}.  
Our notation closely follows Haldane's. \cite{hal81}
We use these results to write $H_+$ and $H_-$ in terms of bosonic operators. This casts
the Hamiltonian $H$ into the form of a spin-boson model with a structured environment.
We also calculate the overlap $\left<F+\right|\left. F-\right>$, where, as stated below Eq.\,\ref{om}, 
$\left|F\pm\right>$ are
the many body ground states of $H_\pm$, which will be relevant when we analyze the
tunneling rate $W$ in Sec.\,\ref{sec5}.  

The free fermion Hamiltonian (\ref{disp}) together with the Fermi sea ground state $\left|F+\right>$
is the starting point for the bosonization procedure.  
Associated with density fluctuations in the
fermion system are operators 
\begin{equation}
a_q=\left(\frac{2\pi}{Lq}\right)^{1/2}\sum_k c_k^\dagger c_{k+q},
\end{equation}
and $a_q^\dagger$, $q=2\pi n/L$, $n=1,\,2,\,3,\,\ldots$ that satisfy
bosonic commutation relations
\begin{equation}
[a_q,a_{q'}]=0,~~[a_q,a_{q'}^\dagger]=\delta_{q,q'}.
\end{equation} 
The bosonic annihilation operators $a_q$ annihilate the Fermi sea $\left|F_+\right>$,
i.e.
\begin{equation}
a_q\left|F+\right>=0.
\end{equation}
A central (and non-trivial) result of bosonization is that, in terms of the bosonic operators,
and for fixed particle number
\begin{equation}
H_+=\sum_{q>0}q\,a_q^\dagger a_q+E_+^{(0)}.\label{hpa}
\end{equation}

The fermion density $\rho(x)$ can be expressed in terms of the bosonic operators as
\begin{equation}
\rho(x)=N/L+\frac{1}{2\pi}\partial_x\left[\varphi(x)+\varphi^\dagger(x)\right],\label{rhotophi}
\end{equation}
where $N=\int_{-L/2}^{L/2}dx\,\rho(x)$ counts the total number of fermions and
\begin{equation}
\varphi(x)=-i\sum_{q>0}\left(\frac{2\pi}{Lq}\right)^{1/2}e^{iqx}a_q.\label{phia}
\end{equation}
The $\varphi$ operators satisfy the commutation relations
\begin{eqnarray}
\left[\varphi(x),\varphi(x')\right]&=&0,\\
\left[\varphi(x),\varphi^\dagger(x')\right]
&=&-\lim_{\eta\to0^+}{\rm ln}\left[1-e^{\frac{i 2\pi}{L}(x-x')-\eta}\right].\label{phicom}
\end{eqnarray}  

Using Eqs.\,\ref{rhotophi} and \ref{phia} to express the potential $V$ in terms
of the bosonic operators, and using expression \ref{hpa} for $H_+$, we find for $H_-$
\begin{align}
&H_-=E_0^{(+)}+\frac{N}{L}v_0\nonumber\\
&+\sum_{q>0}q\left\{a_q^\dagger a_q+\left(\frac{2\pi}{Lq}\right)^{1/2}\left[\frac{v_q^*}{2\pi}a_q
+\frac{v_q}{2\pi}a_q^\dagger\right]\right\},
\end{align} 
where 
\begin{equation}
v_q=\int_{-L/2}^{L/2} dx\,v(x)e^{-iqx}.
\end{equation}
The Hamiltonian $H_-$ is diagonalized by completing the square. For this purpose we define
new bosonic operators 
\begin{equation}
b_q=a_q+\left(\frac{2\pi}{Lq}\right)^{1/2}\frac{v_q}{2\pi},
\end{equation}
that also obey the standard bosonic commutation relations. In terms of these operators
the Hamiltonian $H_-$ reads
\begin{equation}
H_-=E_0^{(+)}+\Delta E +\sum_{q>0} q\,b_q^\dagger b_q,\label{hma}
\end{equation}
where 
\begin{equation}
\Delta E=\frac{Nv_0}{L}-\sum_{q>0}\frac{2\pi}{L}\left|\frac{v_q}{2\pi}\right|^2.\label{deltae}
\end{equation}
Substitution of $H_\pm$ from Eqs.\,\ref{hpa} and \ref{hma} into the full Hamiltonian $H$ of Eq.\,\ref{h1}
reveals that the system is described by the same Hamiltonian as the spin-boson model.
(See for instance Ref.\,\onlinecite{leg87}.)
A quantity that plays a central role in the spin-boson model, is the bosonic environment's spectral function
which in our notation is given by 
\begin{equation}
J(q)=\pi q\left|\frac{v_q}{2\pi}\right|^2.\label{j}
\end{equation}
(Here we have implicitly taken the $L\to\infty$ limit.)
In the context of dissipative quantum mechanics, spectral functions $J(q)\sim q^s$ for $q$ smaller
than some large cut-off, play an important role. The case with $s=1$ is known as an Ohmic environment.
We see that a potential $v(x)$ that is peaked around $x=0$, for instance $v(x)=\lambda/[\pi(x^2+\lambda^2)]$,
produces an Ohmic environment. 
When the spectral function has a more complicated
form, one talks of a structured bath. A
structured bath is obtained by engineering the potential $v(x)$. 
As we shall show in Sec.\,\ref{sec6}, the potential $v(x)$ of Eq. \ref{sinf} produces an environment
with an interesting structure.  
 
The ground state energy of $H_-$ is $E_0^{(+)}+\Delta E$ and the ground state solves $b_q\left|F-\right>=0$, or using
the definition of $b_q$ in terms of $a_q$,
\begin{equation}
a_q\left|F-\right>=-\left(\frac{2\pi}{Lq}\right)^{1/2}\frac{v_q}{2\pi}\left|F-\right>.
\end{equation}
From this follows that the normalized ground state of $H_-$ is the coherent state
\begin{eqnarray}
\left|F-\right>&=&\exp\sum_{q>0}\left(\frac{2\pi}{Lq}\right)^{1/2}\left(\frac{v_q*}{2\pi}a_q-\frac{v_q}{2\pi}a_q^\dagger\right)\left|F_+\right>\nonumber\\
&=&e^{-i\int_{-L/2}^{L/2}dx\,v(x)\left[\varphi(x)+\varphi^\dagger(x)\right]/2\pi}\left|F+\right>.
\label{fmin}
\end{eqnarray}

For future reference we note that the overlap $\left<F+\right|\left.F-\right>$ is easily calculated
from Eq.\,\ref{fmin}. The details of the calculation
can be found in Appendix \ref{appc}.
The result is
\begin{equation}
\left<F+\right|\left.F-\right>=\left(\frac{2\pi}{\Lambda L}\right)^{\alpha/2},
\label{and} 
\end{equation}
where $\Lambda$ is the energy appearing in Eq.\,\ref{cl1}, and, consistent with Eq.\,\ref{alpha0}, (cf. Eq.\,\ref{v0}), 
\begin{equation}
\alpha=(v_0/2\pi)^2.\label{alpha1}
\end{equation} 
For an explicit formula for $\Lambda$,
see Eq.\,\ref{lambda1}. The fact that the overlap tends to zero as $L^{-\alpha/2}$ is known as
the orthogonality catastrophe.\cite{and67} The fact that the same ultraviolet energy $\Lambda$ appears
in the closed loop factor and in the orthogonality catastrophe has previously been established (for
a contact type potential) by Feldkamp and Davis \cite{fel80}, and by Tanaka and Othabe\cite{tan85}. 
   
\section{Closed loop factor}
\label{sec5}
In this section
our goal is to calculate the closed loop factor
\begin{equation}
P(t)=e^{i\Delta E t}\left<F+\right|Q(t)\left|F+\right>,\label{pt}
\end{equation}
where
\begin{equation}
Q(t)=e^{iH_+t}e^{-iH_-t}.
\end{equation}
(The convenience of including the factor $\exp(i\Delta Et)$, with
$\Delta E$ given by Eq.\,\ref{deltae}, will become apparent below.)

Having mapped the system under consideration onto a spin-boson Hamiltonian, we can
simply quote the answer from the literature, namely
\begin{equation}
P(t)=\exp\left[\frac{1}{\pi}\int_{0}^{\infty}dq\, \frac{J(q)}{q^2}\left(e^{-iqt}-1\right)\right],\label{spr}
\end{equation}
with $J(q)$ given by Eq.\,\ref{j}. 
(See for instance Eqs.\,3.35 and 3.36 of Ref.\,\onlinecite{leg87}, but note that $P(t)$ in that work
refers to a different quantity than the one in Eq.\,\ref{pt} of the present work.)  
However, we prefer to give a self-contained derivation
of this result. This derivation goes slightly further than simply calculating $P(t)$, 
namely, it produces a bosonic expression for the operator $Q(t)$ that is normal ordered,
i.e. in which all creation operators are to the left of all annihilation operators.
This expression may in future prove useful for studying non-equilibrium effects. 
Readers prepared to take Eq.\,\ref{spr} as given may wish to skip to the paragraph below
Eq.\,\ref{pinf}.
  
The starting point of our derivation is to consider the time derivative of $Q$, i.e.
\begin{equation}
\partial_t Q(t)=-ie^{iH_+t}Ve^{-iH_+t}Q(t).
\end{equation}
Since the Hamiltonian $H_+$ is also the momentum operator, $\exp(\pm iH_+t)$ is simply translation
by a distance $\pm t$, so  that
\begin{equation}
e^{iH_+t}Ve^{-iH_+t}=\int_{-L/2}^{L/2} dx\,v(x)\rho(x-t)\equiv V(t).\label{bch}
\end{equation}
 Thus we find
\begin{align}
&Q(t)=\mathcal O\exp \left[-i\int_0^tdt' V(t')\right],\label{solq}\\
&=\lim_{n\to\infty}e^{-i\int_{t_{n-1}}^{t_n}dt'\,V(t')}
\times\ldots\times e^{-i\int_{t_{0}}^{t_1}dt'\,V(t')},\label{oexp}
\end{align}
where $t_m=m t/n$, $m=0,\,\ldots,\,n$. 

Let us now consider one of the individual factors in the ordered exponent of Eq.\,\ref{oexp}.
Using Eq.\,\ref{rhotophi} to relate the density operator to the bosonic operators 
$\varphi$ and $\varphi^\dagger$, we find
\begin{align}
&\int_\tau^{\tau+\Delta t}dt'\,V(t')=\int_\tau^{\tau+\Delta t}dt'
\int_{-L/2}^{L/2}dx\,v(x)\nonumber\\
&\hspace{1cm}\times\left\{\frac{N}{L}+\frac{1}{2\pi}\partial_{t'}
\left[\varphi(x-t')+\varphi^\dagger(x-t')\right]\right\}\nonumber\\
&=\Delta tv_0N/L+A(\tau+\Delta t)-A(\tau)
\end{align}
where
\begin{equation}
A(t)=\int_{-L/2}^{L/2}dx\,\frac{v(x)}{2\pi}\left[\varphi(x-t)+\varphi^\dagger(x-t)\right],\label{aop}
\end{equation}
so that
\begin{align}
&e^{-i\int_\tau^{\tau+\Delta t}dt'\,V(t')}=e^{-i\Delta tv_0N/L}e^{-i[A(\tau+\Delta t)-A(\tau)]}.
\end{align}

Since operators $A(\tau)$ and $A(\tau+\Delta t)$ commute to a $c$-number we have
$e^{-i[A(\tau+\Delta t)-A(\tau)]}=e^{-[A(\tau+\Delta t),A(\tau)]/2}e^{-iA(\tau+\Delta t)}e^{iA(\tau)}$. Explicitly
\begin{align}
&[A(\tau+\Delta t),A(\tau)]\nonumber\\
&=-2i\sum_{q>0}\frac{2\pi}{Lq}\left|\frac{v_q}{2\pi}\right|^2\sin[q\Delta t],\label{acom}\\
&=-2i\Delta t\sum_{q>0}\frac{2\pi}{L}\left|\frac{v_q}{2\pi}\right|^2\Delta t+\mathcal O(\Delta t^2).
\end{align}
We therefore find
\begin{equation}
e^{-i\int_\tau^{\tau+\Delta t}dt'\,V(t')}=e^{-i\Delta E\Delta t+\mathcal O(\Delta t^2)}e^{-iA(\tau+\Delta t)}e^{iA(\tau)}.
\end{equation}
When substituted back into Eq.\,\ref{oexp}, this leads to the result
\begin{equation}
Q(t)=e^{-i\Delta E t}e^{-iA(t)}e^{iA(0)}.
\end{equation}  
We can rewrite this as
\begin{equation}
Q(t)=e^{-i\Delta E t}\underbrace{e^{\frac{1}{2}\left[A(t),A(0)\right]}}_{F_1}\underbrace{e^{-i\left[A(t)-A(0)\right]}}_{F_2}.
\label{qt0}
\end{equation}
Factor $F_1$ is what Gutman et al.\cite{gut10} (see their footnote 36) calls the ``anomalous'' 
contribution to the closed loop factor. 
An expression involving the determinant of an operator acting on single particle
Hilbert space often appears in the literature\cite{muz03,amb04,aba04,aba05} 
in connection with the closed loop factor. In Appendix \ref{appe} we show that
this determinant is equal to the expectation value of factor $F_2$. 

Below we consider factors $F_1$ and $F_2$ separately. From Eq.\,\ref{acom} we have
\begin{equation}
e^{\frac{1}{2}\left[A(t),A(0)\right]}=\exp\left\{-i\sum_{q>0}\frac{2\pi}{Lq}\left|\frac{v_q}{2\pi}\right|^2\sin(qt)\right\}.\label{t1}
\end{equation} 
We write factor $F_2$ in boson normal ordered form. This is done to facilitate the calculation of the expectation
value with respect to $\left|F+\right>$.
\begin{eqnarray}
e^{-i\left[A(t)-A(0)\right]}&=&e^{-i\left[B^\dagger(t)-B^\dagger(0)\right]}e^{-i\left[B(t)-B(0)\right]}\nonumber\\
&&\times e^{\frac{1}{2}\left[B^\dagger(t)-B^\dagger(0),B(t)-B(0)\right]}.\label{t2}\\
B(t)&=&\int_{-L/2}^{L/2}dx\,\frac{v(x)}{2\pi}\varphi(x-t).
\end{eqnarray}
Explicitly evaluating the commutator in Eq.\,\ref{t2}, we find
\begin{align}
e^{-i\left[A(t)-A(0)\right]}=&e^{-i\left[B^\dagger(t)-B^\dagger(0)\right]}e^{-i\left[B(t)-B(0)\right]}\nonumber\\
&\times\exp\left\{\sum_{q>0}\frac{2\pi}{Lq}\left|\frac{v_q}{2\pi}\right|^2\left[\cos(qt)-1\right]\right\}.\label{t2b}
\end{align}
Combining this with the result in Eq.\,\ref{t1} for factor $F_1$, we obtain
\begin{align}
Q(t)=&e^{-i\Delta Et}e^{-i\left[B^\dagger(t)-B^\dagger(0)\right]}e^{-i\left[B(t)-B(0)\right]}\nonumber\\
&\times\underbrace{\exp\left\{\sum_{q>0}\frac{2\pi}{Lq}\left|\frac{v_q}{2\pi}\right|^2\left[e^{-iqt}-1\right]\right\}}_{P(t)}.\label{qt}
\end{align}
In the expression for $P(t)$, the infinite system size limit can straight-forwardly be taken to obtain
\begin{equation}
P(t)=\exp\left\{\int_0^\infty dq\,\left|\frac{v_q}{2\pi}\right|^2\frac{e^{-iqt}-1}{q}\right\},\label{pinf}
\end{equation}
in agreement with Eq.\,\ref{spr}.

The large time asymptotics of Eq.\,\ref{pinf} can be extracted as follows. Firstly we write $\ln P(t)$ as  
\begin{eqnarray}
\ln P(t)&=&\int_0^\infty dq\,\left|\frac{v_q}{2\pi}\right|^2\frac{e^{-iqt}-1}{q}\nonumber\\
&=&\lim_{y\to0^+}
\int_0^\infty dq\,\left(\left|\frac{v_q}{2\pi}\right|^2-\left|\frac{v_0}{2\pi}\right|^2
+\left|\frac{v_0}{2\pi}\right|^2\right)\nonumber\\
&&\hspace{1cm}\times\frac{e^{-iqt}-1}{q}e^{-qy}.
\end{eqnarray} 
This expression is then split up into three terms, $\ln P(t)=\lim_{y\to0^+}(T_1+T_2+T_3)$, where
\begin{eqnarray}
T_1&=&\int_0^{\infty} dq\,\left|\frac{v_0}{2\pi}\right|^2e^{-qy}\frac{e^{-iqt}-1}{q},\nonumber\\
T_2&=&-\int_0^\infty dq\, \frac{e^{-qy}}{q}\left(\left|\frac{v_q}{2\pi}\right|^2
-\left|\frac{v_0}{2\pi}\right|^2\right),\nonumber\\
T_3&=&\int_0^\infty dq\, \frac{e^{-q(y-it)}}{q}\left(\left|\frac{v_q}{2\pi}\right|^2
-\left|\frac{v_0}{2\pi}\right|^2\right).\label{terms}
\end{eqnarray}
The integral in $T_1$ is straight forward and leads to $T_1=-\alpha\ln(1+it/y)$ where $\alpha=(v_0/2\pi)^2$
as in Eq.\,\ref{alpha1}, and consistent with Eq.\,\ref{alpha0}.
In Appendix \ref{appd} we show that $T_3=\mathcal O(t^{-1})$ and hence vanishes in the large $t$ limit. Term $T_2$
can be written as
\begin{equation}
T_2=-\alpha\int_0^\infty dq\,\frac{e^{-qy}}{q}\left(\left|\frac{ v_q}{v_0}\right|^2-1\right).
\end{equation}
Thus, for large $|t|$, we obtain $P(t)\simeq (i\Lambda t)^{-\alpha}$ where
\begin{equation}
\Lambda=\lim_{y\to0^+}\left[\frac{1}{y}\exp{\int_0^\infty dq\, \frac{e^{-qy}}{q}\left(\left|\frac{v_q}{v_0}\right|^2-1\right)}\right].
\label{lambda1}
\end{equation}
This implies that $\Lambda$ is determined by the shape of $v(x)$ but not by its overall magnitude: The transformation $v(x)\to c\,v(x)$
does not affect $\Lambda$.

\section{Semi-infinite wire}
\label{sec6}
We apply the results of the previous section to the case of a semi-infinite wire
with $v(x)$ given by Eq.\,\ref{sinf}
so that 
\begin{equation}
v_q=\lambda\cos(ql)e^{-|q|a}.
\end{equation}    
Substitution into Eq.\,\ref{pinf} then yields
\begin{eqnarray}
P(t)&=&C\left\{\left(1+\frac{it}{2a}\right)^2
\left[\left(1+\frac{it}{2a}\right)^2+\left(\frac{l}{a}\right)^2\right]\right\}^{-\alpha/4}\nonumber\\
&=&C\left[1+\frac{it}{2a}\right]^{-\alpha}\left[1+\frac{(l/a)^2}{(1+it/2a)^2}\right]^{-\alpha/4},\label{ptl}
\end{eqnarray}
where $C=\left[1+\left(l/a\right)^2\right]^{\alpha/4}$. Thus in the limit of large $t$,
$P(t)\simeq(i\Lambda t)^{-\alpha}$ where 
\begin{equation}
\Lambda=\left\{2\sqrt{al}\left[1+\left(\frac{a}{l}\right)^2\right]\right\}^{-1}.\label{lam}
\end{equation}
This last result could also have been obtained using Eq.\,\ref{lambda1}.
We see that $P(t)$, and thus also the tunneling rate $W$, grows as $l^{\alpha/2}$ for $l\gg a$,

Some insight into the origin of this result may be obtained by considering
the result in Eq.\,\ref{and} for the overlap $\left<F+\right|\left.F-\right>$, that
displays the orthogonality catastrophe. For $\Lambda$ given by Eq.\,\ref{lam},
$\left|\left<F+\right|\left.F-\right>\right|^2\propto l^{\alpha/2}$.
Therefore increasing $l$
mitigates the orthogonality catastrophe.
The tunneling rate $W$ can be written as (cf. Eq.\,\ref{fgr})
\begin{equation}
W=2\pi|\gamma|^2\sum_n\delta\left(\varepsilon+E_0^{(+)}-E_{n}^{(-)}\right)\left|\left<F+\right|\left.n-\right>\right|^2,\label{fgr2}
\end{equation}
where $E_n^{(-)}$ and $\left|n-\right>$ are the energies and many-body eigenstates of $H_-$, the Hamiltonian
that describes the electrons when the qubit is in state $\left|-\right>$.
The results of Appendix \ref{appc} can be extended to show that not only the ground state to ground state
overlap, but every overlap $\left|\left<F+\right|\left.n-\right>\right|^2$ where $\left|n-\right>$ contains
a finite number of particle hole excitations on top of $\left|F-\right>$, scales like $l^{\alpha/2}$.
Thus the scaling behavior of $W\sim l^{\alpha/2}$ can be understood as a consequence
of the mitigation of the orthogonality catastrophe. The effect relies on the phase coherence of electrons in 
the section of the wire between $|x|=l$ and $x=0$. Hence it is destroyed if the electron phase is randomized
by impurity scattering or if there is inelastic scattering. Thus the increase in $W$ with increasing $l$ 
should persist until $l$ exceeds either the elastic or inelastic mean free path in the wire. Of course the result
is also only valid as long as $\omega\lesssim 1/l$ so that excitations created by the qubit do not resolve the 
spatial structure of the potential. Thus, the larger $l$, the smaller the energy window $0<\omega\lesssim1/l$ in
which the enhancement of $W$ due to mitigation of the orthogonality catastrophe can be observed.

\begin{figure}[tbh]
\begin{center}
\includegraphics[width=\columnwidth]{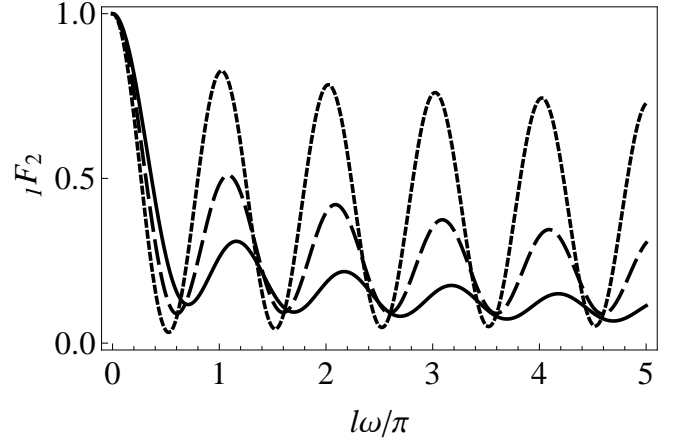}
\caption{The function $_1F_2\left(\frac{\alpha}{4};\frac{\alpha}{2},\frac{\alpha+1}{2};-(l\omega)^2\right)$
that determines the finite $\omega$ behavior of the tunneling rate $W$, as a function of $\omega$ for
various couplings $\alpha$, for the potential $v(x)$ of Eq.\,\ref{sinf}. 
The dotted curve corresponds to $\alpha=1/8$, the dashed curve to $\alpha=1/2$,
and the solid curve to $\alpha=1$. Oscillations with period $\pi/l$ result from the resonant creation of
a single particle hole pair by the qubit. At larger $\alpha$ the oscillations
are washed out due to Fermi-sea shake-up, i.e. the excitation by the qubit of a multitude of particle hole pairs with a
broad distribution of energies.
\label{f1}}
\end{center}
\end{figure}

The tunneling rate $W$ is calculated by expanding the third factor in Eq.\,\ref{ptl} in a Taylor series in
$(1+it/2a)$ and Fourier transforming each term separately. 
Using the identities
\begin{equation}
\int_{-\infty}^\infty dt\, e^{i\omega t}(1+it/r)^{-s}
=\frac{2\pi}{\Gamma(s)}\frac{(r\omega)^s}{\omega}e^{-r\omega},
\end{equation}
for $s,\,t>0$, and 
\begin{equation}
(s)_n\equiv\frac{\Gamma(s+n)}{\Gamma(s)}=\prod_{m=0}^{n-1}(s+m),
\end{equation} 
we obtain
\begin{equation}
W(\omega)=C\,W_0(\omega)
\sum_{n=0}^\infty\frac{\left(\frac{\alpha}{4}\right)_n}{n!(\alpha)_{2n}}\left[-(2l\omega)^2\right]^n,\label{Wl1}
\end{equation}
where $W_0(\omega)$ is the $l=0$ result
\begin{equation}
W_0(\omega)=\frac{2\pi|\gamma|^2}{\Gamma(\alpha)}\frac{(2a\omega)^\alpha}{\omega}e^{-2a\omega}\theta(\omega).
\label{w0}
\end{equation}
The factor $(\alpha)_{2n}$ can be rewritten
\begin{equation}
(\alpha)_{2n}=2^{2n}\left(\frac{\alpha}{2}\right)_n
\left(\frac{\alpha+1}{2}\right)_n.
\end{equation}
Substituting this into Eq.\,\ref{Wl1} we identify the series as the Taylor expansion of the hypergeometric
function $_1F_2$, yielding one of our main results
\begin{equation}
W(\omega)=C\,W_0(\omega)\,
_1F_2\left(\frac{\alpha}{4};\frac{\alpha}{2},\frac{\alpha+1}{2};-(l\omega)^2\right).\label{Wl2}
\end{equation}
As $\omega\to 0^+$, $_1F_2\left(\frac{\alpha}{4};\frac{\alpha}{2},\frac{\alpha+1}{2};-(l\omega)^2\right)$
tends to $1$, so that $W(\omega)$ indeed has the expected power law singularity for small $\omega$ (cf. Eq.\,\ref{w}) with 
$\Lambda$ given by Eq.\,\ref{lam}. When $\omega$ becomes of the order $1/l$, excitations in the wire 
are able to resolve the spatial structure of the potential $v(x)$ and $W(\omega)$ starts deviating from 
simple power law behavior. In the weak coupling limit, i.e. small $\alpha$, the hypergeometric function
$_1F_2\left(\frac{\alpha}{4};\frac{\alpha}{2},\frac{\alpha+1}{2};-(l\omega)^2\right)$
reduces to $\cos(l\omega)^2$, and $W$ therefore shows oscillations with period $\pi/l$ as a function of
$\omega$. These oscillations can be understood as being due to the resonant
creation of a single particle-hole excitation with an energy that satisfies the resonance condition $\omega=\pi n/l$.
(This condition maximizes the single particle matrix element $\left<h\right|V\left|p\right>$, where
$h$ and $p$ refer to the single particle orbitals of the hole and the excited particle respectively.)

As shown in Figure \ref{f1}, the oscillations become damped as $\alpha$ is increased. The damping is a signature of
a phenomenon known as Fermi-sea shake-up.
At strong coupling (large $\alpha$), rather than creating a single particle hole pair, 
a large number of particle hole pairs are created. This corresponds
to many charge density excitations (created by the bosonic operators $a_q^\dagger$) with a broad
distribution of wavelengths $1/q$. As a result there is no clear resonance any more, and the
oscillations in $W(\omega)$ are washed out. 

It is also instructive to investigating the regime
$1/l\ll\omega$. Here the asymptotic behavior of the hypergeometric function is
\begin{equation}
_1F_2\left(\frac{\alpha}{4};\frac{\alpha}{2},\frac{\alpha+1}{2};-(l\omega)^2\right)\simeq\frac{\Gamma(\alpha)}{\Gamma(\alpha/2)}(2l\omega)^{-\alpha/2}.
\end{equation}
If we substitute this into the expression (Eq.\,\ref{Wl2}) for $W$, assuming $l\gg a$, we obtain
\begin{equation}
\left.W(\omega)\right|_{\omega\gg1/l}=\frac{2\pi|\gamma|^2}{\Gamma(\alpha/2)}\frac{(2a\omega)^{\alpha/2}}{\omega}e^{-2a\omega}\theta(\omega).
\label{wbigw}
\end{equation}
This is the same rate as would be obtained from a closed loop factor
\begin{equation}
P(t)=\left[\left(1+\frac{it}{2a}\right)^{\left(v_0/4\pi\right)^2}\right]^2.
\end{equation}
Such a closed loop factor could also be obtained by coupling the qubit to two independent chiral channels,
where the potential the qubit produces in either channel equals $u(x)$ of Eq.\,\ref{sinf}, i.e. one of the
peaks in the full potential $v(x)=u(x+l)+u(x-l)$. This means that, in the single channel semi-infinite wire, 
at energies $\omega\gg1/l$, the potential $u(x-l)$ experienced by right-moving electrons  
and the potential $u(x+l)$ experienced by left-moving electrons, contribute incoherently to the rate
$W$, as if left-movers and right-movers belong to separate channels. 
Could this indicate that  
electrons reflected at the tip of the wire have lost all memory of their in-bound encounter with the qubit 
by the time that they
again reach the qubit on their out-bound journey? Since the
electrons undergo no relaxation between encounters with the qubit, the answer is ``no''. 
Rather, what Eq.\,\ref{wbigw} indicates, is that processes in which 
an individual electron wave-packet with width $1/\omega\ll l$ is scattered twice, once while incident on the tip,
and once after being reflected at the tip, are rare and make a vanishingly small contribution to the rate $W(\omega)$.   

As stated in the introduction, $W$ corresponds to the exponential decay rate of the probability to
find the qubit in state $\left|+\right>$, provided that $1\gg W(\omega)/\omega$.  
We conclude this section by investigating when this inequality holds. For $\alpha<2$, $W/\omega$ diverges when $\omega\to0^+$,
and the criterion for exponential decay is violated at small $\omega$. 
From the small $\omega$ asymptotics $W(\omega)/\omega\sim |\gamma|^2 (\omega/\Lambda)^\alpha/\omega^2$, we
conclude that, for $\alpha<2$, exponential decay with rate $W$ occurs when
\begin{equation}
\omega\gg |\gamma| \left|\gamma/\Lambda\right|^{\alpha/(2-\alpha)},\label{e1}
\end{equation}
When $\alpha>2$, on the other hand, $W(\omega)/\omega$ no longer diverges, but rather reaches
a maximum value of order $|a\gamma|^2$ at $\omega\sim1/a$. Thus, for $\alpha>2$, exponential decay with
rate $W$ occurs for all $\omega>0$, provided that
\begin{equation}
1\gg|a\gamma|^2.\label{e2}
\end{equation}
The same regime for exponential decay as in Eqs.\,\ref{e1} and \ref{e2} was identified more rigorously by Legget et al.\cite{leg87} 
in the context of the spin boson Hamiltonian with an unstructured Ohmic bath (corresponding to $l=0$ in our system).
(See their Sec. VII.B, and in particular their Eq.\,7.17a. Note that the quantity that we denote $\alpha$ is
twice the quantity that they denote $\alpha$.) The fact that our qubit is immersed in a structured
bath does not affect the result because the bath spectral function $J(q)$ still has the same large and
small $q$ asyptotics as in the case of an Ohmic bath.
 
\section{Summary and conclusions}
\label{sec3}
In this paper we studied the quantity $\left<F+\right|e^{iH_+t}e^{-iH_-t}\left|F+\right>$,
known in the language of the Fermi edge singularity as closed loop factor, for the case where
the Hamiltonians $H_\pm$ describe electrons in a single chiral channel. 
We investigated the regime $E_F\gg v_F/a$ where
$a$ is the typical length scale on which the potentials $V_\pm$ associated with
$H_\pm$ vary, a regime not studied before.  We investigated a system
where the Fourier transform of the closed loop factor gives the tunneling rate of a two-level system
(charge qubit) coupled to the electron gas.   
Under the assumption of linear dispersion 
(cf. Eq.\,\ref{disp}) we obtained the exact expression 
for the closed loop factor  
valid for arbitrary $V_\pm$ and arbitrary times. We studied its large time asymptotics, and
obtained an exact formula for the ultraviolet energy $\Lambda$ that appears in Eq.\,\ref{cl1}.
Unlike in the previously studied regime where $\Lambda\sim E_F$, here we find $\Lambda<1/a\ll E_F$.
Furthermore it turns out that $\Lambda$ is determined by the shape, but not the overall magnitude 
of the potentials $V_\pm$, i.e. scaling $V_\pm\to c\,V_\pm$ 
leaves $\Lambda$ invariant. 

We applied our general results to the example       
a semi-infinite wire. The qubit interacts with the wire at a point that is a distance $l$ from
the tip of the wire. In this system
we found that the tunneling rate $W$ could be enhanced without increasing either 
$\gamma$ or $\alpha$. $\Lambda$ decreases like $l^{-1/2}$ so that 
$W$ grows like $l^{\alpha/2}$.  Thus the tunneling rate $W$
becomes larger the further the qubit is from the tip of the wire. 
This effect is due to a mitigation of the orthogonality 
catastrophe. It holds as long as $l$ is
less than the phase-coherence length of electrons in the wire
and for level splittings $0<\omega\lesssim 1/l$.

Armed with an expression for the closed loop factor that
is valid also for small times, we obtain an exact expression for $W$ for the
semi-infinite wire system. 
The finite $\omega$ features of $W$ probe the 
spatial profile of the potential $V_\pm$ at length scales $1/\omega$ (in units where $v_F=1$).
We study how the $\omega$ dependence of $W$ changes as the coupling $\alpha$
between the qubit and the electron gas grows.
At weak coupling (small $\alpha$), we find that the rate $W$ oscillates as a function of $\omega$, 
and that the period is $\pi/l$.
This is due to the resonant excitation of a single particle-hole pair in the Fermi sea.
The wavelength of the associated charge density fluctuation is $\pi/\omega$. The resonance condition
is that an integer number of half wavelengths fit into the part of the wire between the tip and 
the point where the qubit interacts with the wire. 
At strong coupling (large $\alpha$) on the other hand, many particle hole excitations are created. 
This is known as Fermi sea shake-up.
The corresponding density fluctuations have a broad distribution of wavelengths
and hence there are no clear resonances. This results in the damping of the oscillations in $W(\omega)$
as $\alpha$ is increased. One of our main results (Eq.\,\ref{Wl2}) is an exact formula for this damping by
means of Fermi sea shake-up. The result is illustrated in Fig.\,\ref{f1}. 
We also analyzed rate $W$ in the $\omega\gg 1/l$ limit
and saw that here the left and right moving electrons
contribute to the rate $W$ as if they belong to independent channels.  
This happens despite the fact that each electron incident on the tip encounters the qubit twice, once before
being reflected at the tip, and once afterwards, and no electron relaxation occurs between qubit encounters.  
The result therefore indicates that processes in which 
an individual electron wave-packet with width $1/\omega\ll l$ is scattered twice, once while incident on the tip
and once after being reflected at the tip, make a vanishingly small contribution to the rate $W(\omega)$.   

\appendix

\section{Obtaining $W$ from Fermi's golden rule}
\label{appa}
In this Appendix we apply Fermi's golden rule to obtain the expression in Eq.\,\ref{w1} for the
transition rate $W$. The initial state for the transition is 
$\left|F+\right>\otimes\left|+\right>$ with energy $E_0^{(+)}+\varepsilon$.
Possible final states are of the form $\left|n-\right>\otimes\left|-\right>$, where $\left|n-\right>$ is an eigenstate
of $H_-$ and has energy $E_n^{(-)}$. We have to sum over all eigenstates of $H_-$. Thus
\begin{eqnarray}
W_{-+}&=&2\pi|\gamma|^2\sum_n\delta\left(\varepsilon+E_0^{(+)}-E_n^{(-)}\right)\left|\left<F+\right|\left.n-\right>\right|^2\nonumber\\
&=&|\gamma|^2\sum_n\int_{-\infty}^\infty dt\,e^{i\left(\varepsilon+E_0^{(+)}-E_n^{(-)}\right)t}\left|\left<F+\right|\left.n-\right>\right|^2\nonumber\\
&=&|\gamma|^2\sum_n\int_{-\infty}^\infty dt\,e^{i\varepsilon t}\left<F+\right|e^{iH_+t}\nonumber\\
&&\hspace{2cm}\times\left|n-\right>\left<n-\right|e^{-iH_-t}\left|F+\right>\nonumber\\
&=&|\gamma|^2\int_{-\infty}^\infty dt\,e^{i\varepsilon t}\left<F+\right|e^{iH_+t}e^{-iH_-t}\left|F+\right>.\label{fgr}
\end{eqnarray} 

\section{Gauging away $V_+$.}
\label{appaa}
In the main text we chose the potential $V_+$ as zero, and stated that this does not
involve any loss of generality. Here we prove this claim. Suppose 
\begin{equation}
H_\pm=\int_{-L/2}^{L/2}dx\,\psi^\dagger(x)\left(-i\partial_x-\mu+v_\pm(x)\right)\psi(x),
\end{equation}
Now define position dependent phases
\begin{equation}
\lambda_\pm(x)=-\int_{-L/2}^xdx'\,v_\pm(x'),
\end{equation}
and total phase shifts $\lambda_\pm=\lambda_\pm(L/2)$. Then define a new set of fermion operators
$\bar\psi(x)$ related to $\psi(x)$ by
\begin{equation}
\bar\psi(x)=e^{i\left[\lambda_+(x)-\lambda_+x/L\right]}\psi(x).
\end{equation}
The operators $\bar\psi^\dagger(x')$ and $\bar\psi(x)$ obey the same anti-commutation relations as 
$\bar\psi^\dagger(x')$ and $\bar\psi(x)$ and are also periodic with period $L$.

In terms of $\bar\psi(x)$ and $\bar\psi^\dagger(x)$, the Hamiltonian $H_+$ has the form
\begin{equation}
H_+=\int_{-L/2}^{L/2}dx\,\bar\psi^\dagger(x)\left(-i\partial_x-\mu-\lambda_+/L\right)\bar\psi(x),\label{hp}
\end{equation}
while 
\begin{align}
H_+=\int_{-L/2}^{L/2}dx\,\bar\psi^\dagger(x)\big(&-i\partial_x-\mu-\lambda_+/L\nonumber\\
&+v_-(x)-v_+(x)\big)\bar\psi(x),\label{hm}
\end{align}
Thus, in terms of the new fermion operators, $H_+$ and $H_-$ are of the same form as in Eqs.\,\ref{disp} and
\ref{pot}, with $v(x)\to v_-(x)-v_+(x)$ and $\mu\to\mu+\lambda_+/L$.

\section{Anderson's orthogonality catastrophe}
\label{appc}
Anderson\cite{and67} states that the overlap $\left<F+\right|\left.F-\right>$ vanishes as a power law $L^{-\alpha/2}$
as the system size grows. For the present system we can calculate this overlap exactly
for arbitrary potentials $v(x)$. Our starting point is Eq.\,\ref{fmin} and the operator identity
$e^{A+B}=e^Ae^Be^{-[A,B]/2}$, provided that $[A,[A,B]]=[B,[A,B]]=0$. 
\begin{eqnarray}
&&\left<F+\right|\left.F-\right>\nonumber\\
&=&\left<F+\right|\exp\sum_{q>0}\left(\frac{2\pi}{Lq}\right)^{1/2}\left(\frac{v_q}{2\pi}a_q-\frac{v_q*}{2\pi}a_q^\dagger\right)\left|F+\right>\nonumber\\
&=&\exp\left\{-\frac{1}{2}\sum_{q>0}\frac{2\pi}{Lq}\left|\frac{v_q}{2\pi}\right|^2\right\}.\label{and1}
\end{eqnarray}
In the large $L$ limit, we rewrite this as 
\begin{align}
&\left<F+\right|\left.F-\right>\nonumber\\
&=\lim_{y\to0^+}\Bigg[\underbrace{\frac{1}{y}\exp\left[\int_0^\infty dq\,e^{-qy}\left(\left|\frac{v_q}{v_0}\right|^2-1\right)\right]}_1\nonumber\\
&\hspace{2cm}\times \underbrace{y\exp\left(\sum_{q}^\infty \frac{2\pi}{Lq}e^{-qy}\right)}_2\Bigg]^{-\alpha/2}, 
\end{align}
where $\alpha=|v_0/2\pi|^2$ as in Eq.\,\ref{alpha1}. The $y\to0^+$ limit of the two factors marked $1$ and $2$ can be
taken separately. Referring back to Eq.\,\ref{lambda1}, we identify  
the factor marked $1$ as the energy $\Lambda$ that appears in Eqs.\,\ref{cl1}. The sum in the exponent of the factor marked $2$
is the Taylor expansion of the logarithm function and hence
\begin{align}
y\exp\left(\sum_{q}^\infty \frac{2\pi}{Lq}e^{-qy}\right)&=y\left[1-\exp(-2\pi y/L)\right]^{-1}\nonumber\\
&\simeq L/2\pi.
\end{align}
This leads to the result
\begin{equation}
\left<F+\right|\left.F-\right>=\left(\frac{2\pi}{\Lambda L}\right)^{\alpha/2}.
\end{equation}

\section{Asymptotics of $P(t)$}
\label{appd}
In the main text, in the derivation of the asymptotic form of $P(t)$, we stated
that $T_3$ in Eq.\,\ref{terms} vanishes like $1/t$ in the large $|t|$ limit. Here we give a proof.
By Fourier transforming from $v_q$ to $v(x)$ we obtain
\begin{align}
T_3=\int dx\int dx'&\frac{v(x)}{2\pi}\frac{v(x')}{2\pi}\nonumber\\
&\underbrace{\int_0^\infty dq\,\frac{e^{-iq(t-iy)}}{q}\left[e^{iq(x-x')}-1\right]}_I.
\end{align}
The integral $I$ can be performed to obtain
\begin{equation}
I=\ln\left(1+\frac{it}{y}\right)-\ln\left(1+\frac{i(t-x+x')}{y}\right).
\end{equation}
Expanding in $1/t$ we find $I=(x-x')/t+\mathcal O(t^{-2})$.

\section{Determinantal formula related to closed loop contribution}
\label{appe}

Here we show that the expectation value of the factor $F_2$ in 
Eq.\,\ref{qt0} with respect to $\left|F+\right>$ equals a determinant of an operator acting on single particle
Hilbert space. 

The proof relies on the following general result for fermionic systems.
Let $B=\{\left|n\right>| n=1,\,2,\,\ldots\}$ be a set of orthonormal single particle orbitals
and let $c_n^\dagger$ and $c_n$ be the associated fermionic creation and annihilation operators.
Let $F$ be a subset of $B$. Without loss of generality, we may take 
$F=\{\left|n\right>| n=1,\,2,\,\ldots\,N\}$.
Let $\left|F\right>$ be the many-fermion state
\begin{equation}
\left|F\right>=\prod_{m=1}^Nc^\dagger_m\left|0\right>.
\end{equation}
Let $H$ be the operator 
\begin{equation}
H=\sum_{m,n=1}^{\infty} h_{mn}c_m^\dagger c_n.
\end{equation}
Then $e^{iH}\left|F\right>=\prod_{m=1}^N\tilde c^\dagger_m\left|0\right>$,
where the fermionic operator $\tilde c^\dagger_m$ creates a particle in the 
orbital $\left|\tilde n\right>=e^{ih}\left|n\right>$, where
\begin{equation}
h=\sum_{m,n=1}^{\infty} h_{mn}\left|m\right>\left<n\right|,
\end{equation}
is an operator acting on single particle Hilbert space.
This implies that $\left<F\right|e^{iH}\left|F\right>=\det e^{ih}_{FF}$ is a Slater determinant
where $e^{ih}_{FF}$ is an $N\times N$ matrix with entries
\begin{equation}
\left[e^{ih}_{FF}\right]_{mn}=\left<m\right|e^{ih}\left|n\right>.
\end{equation}
The operator $f=\sum_{n=1}^N\left|n\right>\left<n\right|$ projects onto the subspace spanned by the set $F$,
and hence the matrix $e^{ih}_{FF}$ has the same determinant as the operator $1-f+e^{ih}f$, leading to
the result
\begin{equation}
\left<F\right|e^{iH}\left|F\right>=\det\left(1-f+e^{ih}f\right).\label{det}
\end{equation}
We derived this result for a state $\left|F\right>$ containing a finite number of particles.
We postulate that a similar result holds for the state $\left|F+\right>$ representing an 
infinitely deep Fermi sea.
 
In order to apply the above result to $F_2=e^{-i[A(t)-A(0)]}$ we have to show that $A(t)-A(0)$
is quadratic in fermion creation and annihilation operators. This is indeed so, as is seen by referring to 
Eq.\,\ref{rhotophi} and Eq.\ref{aop} to obtain
\begin{equation}
A(t)-A(0)=\int_0^t dt' \int_{-L/2}^{L/2} dx\,\left[v(x-t')-\frac{v_0}{L}\right]\psi(x)^\dagger\psi(x).
\end{equation}
Thus we may use Eq.\,\ref{det} to write
\begin{equation}
\left<F+\right|e^{-i[A(t)-A(0)]}\left|F+\right>=\det\left(1-n +e^{-i\delta}n\right),
\end{equation}
where $n$ and $\delta$ are the single particle operators
\begin{eqnarray}
n&=&\sum_{k<\mu}\left|k\right>\left<k\right|,\nonumber\\
\delta&=&\int_0^t dt'\int_{-L/2}^{L/2}\left[v(x-t')-\frac{v_0}{L}\right]\left|x\right>\left<x\right|.
\end{eqnarray}  

\acknowledgments

This research was supported by the National Research Foundation (NRF) of South Africa.


\begin{thebibliography}{99}
\bibitem{kon64} J.\,Kondo, Prog. Theor. Phys. {\bf 32}, 37, (1964).
\bibitem{mah67} G.\,D.\,Mahan, Phys. Rev. {\bf 163}, 612, (1967).
\bibitem{noz69} P.\,Nozi\`{e}res and C.\,T.\,DeDominicis, Phys. Rev. {\bf 178}, 1097, (1969).
\bibitem{pus04} M. Pustilnik and L. I. Glazman, J. Phys.: Condens. Matter {\bf 16}, R513, (2004).
\bibitem{aba04} D.\,A.Abanin and L.\,S.\,Levitov, Phys. Rev. Lett. {\bf 93}, 126802, (2004).
\bibitem{muz03} B.\,Muzykantskii, N.\,d'Ambrumenil,\, and B.\,Braunecker, Phys. Rev. Lett. {\bf 91}, 266602, (2003).
\bibitem{amb04} N.\,d'Ambrumenil and B.\,Muzykantskii, arXiv:cond-mat/0405457.
\bibitem{aba05} D.\,A.\,Abanin and L.\,S.\,Levitov, Phys. Rev. Lett. {\bf 94}, 186803, (2005).
\bibitem{sny07} I. Snyman and Yu. V. Nazarov, Phys. Rev. Lett. {\bf 99}, 096802, (2007).
\bibitem{gut10} D.\,B.\,Gutman, Y.\,Gefen, and A.\,D.\,Mirlin, Phys. Rev. B {\bf 81}, 085436, (2010).
\bibitem{bet11} E. Bettelheim, Y. Kaplan, and P. Wiegmann, J. Phys. A: Math. Theor. {\bf 44}, 282001, (2011).
\bibitem{gut11} D.\,B.\,Gutman, Y.\,Gefen, and A.\,D.\,Mirlin, J. Phys. A: Math. Theor. {\bf 44}, 165003, (2011).
\bibitem{bet11b} E.\,Bettelheim, Y.\,Kaplan, and P.\,Wiegmann, Phys. Rev. Lett. {\bf 106}, 166804, (2011).
\bibitem{mkh11} V.\,V.\,Mkhitaryan and M.\,E.\,Raikh, Phys Rev. Lett. {\bf 106}, 197003, (2011).
\bibitem{elz03} J.\,M.\,Elzerman, R.\,Hanson, J.\,S.\,Greidanus,\,L.\,H.\,Willems van Beveren, S. De Franceschi, 
L.\,M.\,K.\,Vandersypen, S.\,Tarucha, and L.\,P.\,Kouwenhoven, Phys. Rev. B {\bf 67}, 161308R, (2003).
\bibitem{pet04} J.\,R.\,Petta, A.\,C.\,Johnson, C.\,M.\,Marcus, M.\,P.\,Hanson, and A.\,C.\,Gossard, Phys. Rev. Lett. {\bf 93}, 186802 (2004).
\bibitem{f1} The exponential decay of $n_+(t)$ can be justified 
using the general textbook argument, cf. Chapter 18 of E. Merzbacher
{\em Quantum Mechanics, 2nd ed.}, (Wiley, New York, 1970). 
Alternatively, exponential decay can be proven
for the specific system that we study by mapping it onto a spin boson Hamiltonian, as we do in Sec.\,\ref{sec4},
and then using the results obtained by Legget et al.,\cite{leg87} by means of their ``non-interacting blip'' approximation.
The advantage of this approach is that it yields a rigorous criterion for the regime of exponential decay. (See their Eq. 7.13.)
It turns out that they find exponential decay in the same region of parameter space as
that predicted by the intuitive argument we provide in the Introduction. See also the discussion at the end of Sec.\,\ref{sec6}.
\bibitem{oth90} K.\,Ohtaka and Y.\,Tanabe, Rev. Mod. Phys. {\bf 62}, 929, (1990).
\bibitem{tan85} Y.\,Tanabe and K.\,Ohtaka, Phys. Rev. B, {\bf 32}, 2036, (1985).
\bibitem{sch59} J.\,Schwinger, Phys. Rev. Lett. {\bf 3}, 296, (1959).
\bibitem{mat65} D.\,C.\,Mattis and E.\,H.\,Lieb, J. Math. Phys. {\bf 6}, 304, (1965).
\bibitem{hal81} F.\,D.\,M.\,Haldane, J.\,Phys.\,C: Solid State Phys. {\bf 14}, 2582, (1981).
\bibitem{del98} J.\,von Delft and H.\,Schoeller, Ann. Phys. {\bf 7}, 225, (1998).
\bibitem{sch69} K.\,D.\,Schotte and U.\,Schotte, Phys. Rev. {\bf 182}, 479, (1969).
\bibitem{leg87} A.\,J.\,Legget, S.\,Chakravarty, A.\,T.\,Dorsey, M.\,P.\,A.\,Fisher, A.\,Garg, and W.\,Zwerger, 
Rev. Mod. Phys. {\bf 59}, 1, (1987).
\bibitem{fab95} M.\,Fabrizio and A.\,O.\,Gogolin, Phys. Rev. B, {\bf 51}, 17827 (1995)
\bibitem{and67} P.\,W.\,Anderson, Phys. Rev. Lett. {\bf 18}, 1049, (1967). 
\bibitem{fel80} L.\,A.\,Feldkamp and L.\,C.\,Davis, Phys. Rev. B {\bf 22}, 4994, (1980).
\end{thebibliography}
\end{document}